\documentclass{article}
\usepackage[T1]{fontenc}
\usepackage{spconfa4,amsmath,graphicx}
\usepackage[nolist]{acronym}
\usepackage{adjustbox}
\usepackage[breaklinks]{hyperref}
\usepackage{colortbl}
\usepackage{orcidlink}
\usepackage{amsfonts,amssymb,amsthm}
\usepackage{booktabs, siunitx}
\usepackage{comment}
\usepackage{multirow}

\begin{acronym}
\acro{MAE}{mean absolute error}
\acro{CNN}{convolutional neural network}
\acro{CRNN}{convolutional recurrent neural network}
\acro{DoA}{direction of arrival}
\acro{DNN}{deep neural network}
\acro{DRR}{direct-to-reverberant ratio}
\acro{ELU}{exponential linear unit}
\acro{GMM}{Gaussian mixture model}
\acro{GRU}{gated recurrent unit}
\acro{IID}{interaural intensity difference}
\acro{ITD}{interaural time difference}
\acro{LSTM}{long short-term memory}
\acro{LPC}{linear predictive coding}
\acro{MSE}{mean squared error}
\acro{RIR}{room impulse response}
\acro{RNN}{recurrent neural network}
\acro{SELD}{sound event localization and detection}
\acro{SNR}{signal-to-noise ratio}
\acro{SVM}{support vector machines}
\acro{STFT}{short-time fourier transform}
\end{acronym}

\title{Dependence on Early and Late Reverberation of Single-Channel Speaker Distance Estimation}

\name{Michael Neri~\orcidlink{0000-0002-6212-9139}, Archontis Politis~\orcidlink{0000-0002-0595-2356}, Tuomas Virtanen~\orcidlink{0000-0002-4604-9729}}
\address{Faculty of Information Technology and Communication Sciences, Tampere University, Finland \\
\textit{\{michael.neri, archontis.politis, tuomas.virtanen\}@tuni.fi}}

\definecolor{Gray}{gray}{0.85}
\definecolor{red_cool}{rgb}{0.5, 0.0, 0.0}

\begin{document}
\ninept
\maketitle
\begin{abstract}
Single-channel speaker distance estimation has recently achieved centimeter-level accuracy in simulated environments, yet it remains unclear which components of the room impulse response (RIR) the model exploits and how performance depends on the recording conditions. In this work, we decompose simulated RIRs into four variants (full, direct-only, no-late, and no-early) using the mixing time estimated from the echo density function as the boundary between early reflections and late reverberation. We define four calibration scenarios, from fully calibrated (synchronised capture, known source level) to fully uncalibrated (arbitrary onset, unknown level), and evaluate all combinations on a matched dataset. Results show that without time calibration, mean absolute error (MAE) increases to $1.29$\,m and the model extracts reverberation-based cues, with early reflections emerging as the most informative component. Further analysis against DRR, $C_{50}$, and $T_{60}$ confirms that estimation accuracy improves with stronger early energy and degrades in highly reverberant environments. When time calibration is available, the model achieves a MAE of $0.14$\,m by extracting the propagation delay alone, regardless of the RIR content.
\end{abstract}
\section{Introduction}
\label{sec:intro}
Estimating the distance of a sound source from a single-channel recording is a problem of practical interest in hearing aid devices~\cite{Hamacher_2005_EURASIP}, hands-free communication~\cite{Oh_1992_ICASSP}, and speech recognition~\cite{Omologo_1198_SpeechComm}.  Unlike \ac{DoA} estimation, which humans solve reliably through binaural cues such as \ac{ITD} and \ac{IID}~\cite{Dietz_2011_SpeechCommunication}, distance perception is inherently more difficult: binaural cues carry negligible distance information beyond close proximity~\cite{zahorik2005auditory}, and listeners must instead rely on a combination of level attenuation, reverberation, and familiarity with the source signal~\cite{zahorik2005auditory, mendoncca2016modeling}.  When these cues are degraded, human distance perception deteriorates considerably~\cite{cherry_1953_JournalAcousAme, Georganti_TASLP_2011}.

Early approaches to source distance estimation operated on binaural recordings and relied on hand-crafted features derived from the \ac{DRR}~\cite{Lu_2010_TASLP} or from inter-channel correlations. Vesa trained \acp{GMM} on binaural correlation features to classify distance into discrete bins~\cite{Vesa_2007_WASPAA, Vesa_2009_TASLP}, while in~\cite{Georganti_2013_TASLP} the standard deviation of inter-channel differences is used with ensembles of \acp{GMM} and \acp{SVM}. These methods typically required careful hyperparameter tuning and generalised poorly across rooms with different acoustic conditions. More recent studies have revisited the problem with \acp{DNN}, yet most remain limited to binary far/near classification~\cite{Patterson_2022_Interspeech, krause2021joint} or to coarse distance bins over a narrow range~\cite{Georganti_TASLP_2011, Yiwere_2019_Sensor, Sobhdel_ICMLA_2024}. Single-channel setups, motivated by the constraints of low-power devices with limited computational resources~\cite{Georganti_TASLP_2011, Venkatesan_2020_CirSys}, have received comparatively little attention. The work in~\cite{Neri_WASPAA_2023} was the first to frame speaker distance as a continuous regression problem, demonstrating that a \ac{DNN} operating on \ac{STFT} phase features can achieve centimeter-level accuracy in simulated environments. It is worth noting that the centimeter-level accuracy reported in~\cite{Neri_TASLP_2024, Neri_WASPAA_2023} was obtained on simulated data where the convolved signal preserves the propagation delay as a leading silence proportional to distance. In practice, however, the signal onset has no fixed relation to the source-microphone distance, making this cue a simulation artifact rather than a usable feature. In addition, that study treated the model as a black box and did not investigate which acoustic cues it leverages. Then, in~\cite{GenDA2025_RoomAcoustics} the authors proposed a data augmentation approach for generating realistic room-acoustic conditions to support speaker distance estimation using the \ac{CRNN} model in~\cite{Neri_TASLP_2024}.

\begin{figure}[t!]
    \centering
    \centerline{\includegraphics[width=0.53\linewidth]{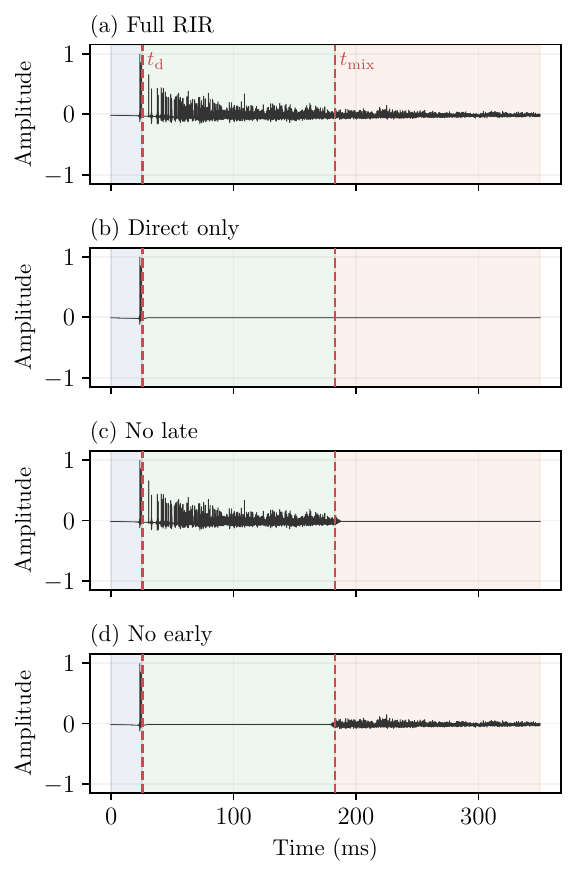}}
    \caption{\ac{RIR} decomposition with source-microphone distance of $d = 7.2$ meters, $t_d = 25.4$ milliseconds and $t_{\mathrm{mix}} = 182.9$ milliseconds.}
    \label{fig:rir_variants}
\end{figure}

These findings raise a natural question that remains unanswered: \emph{which temporal components of the \ac{RIR} does the model actually exploit, and to what extent does each component contribute to estimation accuracy?} A \ac{RIR} between a source and a microphone can be decomposed into three successive regions: the direct path, early reflections, and late reverberation, each carrying qualitatively different information about the source-microphone geometry and the enclosing room. The direct path encodes distance through propagation delay and $1/r$ attenuation, with $r$ being the source-receiver distance; early reflections add geometric constraints imposed by nearby wall surfaces; late reverberation provides a statistical fingerprint of the room's volume and absorption. Understanding which of these regions drives a learned estimator is essential both for interpreting the model's behaviour and for guiding the design of more robust systems.
 
In this work, we present a systematic analysis of the single-channel speaker distance estimation task under controlled calibration conditions. Our contributions are as follows:
\begin{itemize}
  \item We decompose simulated \acp{RIR} into four variants (full, direct-only, no-late, and no-early) using the mean-free-path mixing time~\cite{kuttruff2016room, abel2006} as the boundary between early reflections and late reverberation, and train separate models on each variant to isolate the contribution of every temporal region.
 
  \item We modeled two distance cues beyond reverberation. The first is the propagation delay preserved in simulated data. The second is the source emission level, which in some practical deployments (e.g., a loudspeaker at fixed output) provides a usable amplitude cue via the $1/r$ law. Crossing onset randomisation with level variation on a matched dataset yields four conditions that isolate the contribution of each cue.
 
  \item We present a $4 \times 4$ evaluation (four calibration conditions with four \ac{RIR} variants) that reveals how much of the estimation accuracy is attributable to calibration cues versus reverberation-based acoustic features, and which \ac{RIR} components compensate when calibration information is unavailable.
\end{itemize}

The work is structured as follows: Section~\ref{sec:analysis} provides all the details regarding the generated dataset (including all \ac{RIR} variant and calibration scenarios) and the employed baseline model; Section~\ref{sec:exp} depicts all the results in all conditions whereas Section~\ref{sec:concl} draws the conclusions.

\section{Experimental setup}
\label{sec:analysis}
We consider a single omnidirectional microphone placed at an unknown position $\mathbf{m} \in \mathbb{R}^3$ inside a room, capturing the speech signal of a talker located at an unknown position $\mathbf{s} \in \mathbb{R}^3$ in the same unknown room. Neither the room geometry, the acoustic properties, nor the absolute positions of the source and microphone are assumed to be known. The goal is to estimate the distance $r = \lVert \mathbf{s} - \mathbf{m} \rVert$ from the recorded single-channel signal alone. 

\subsection{Dataset}
\begin{figure*}[t!]
    \centering
    \centerline{\includegraphics[width=0.81\linewidth]{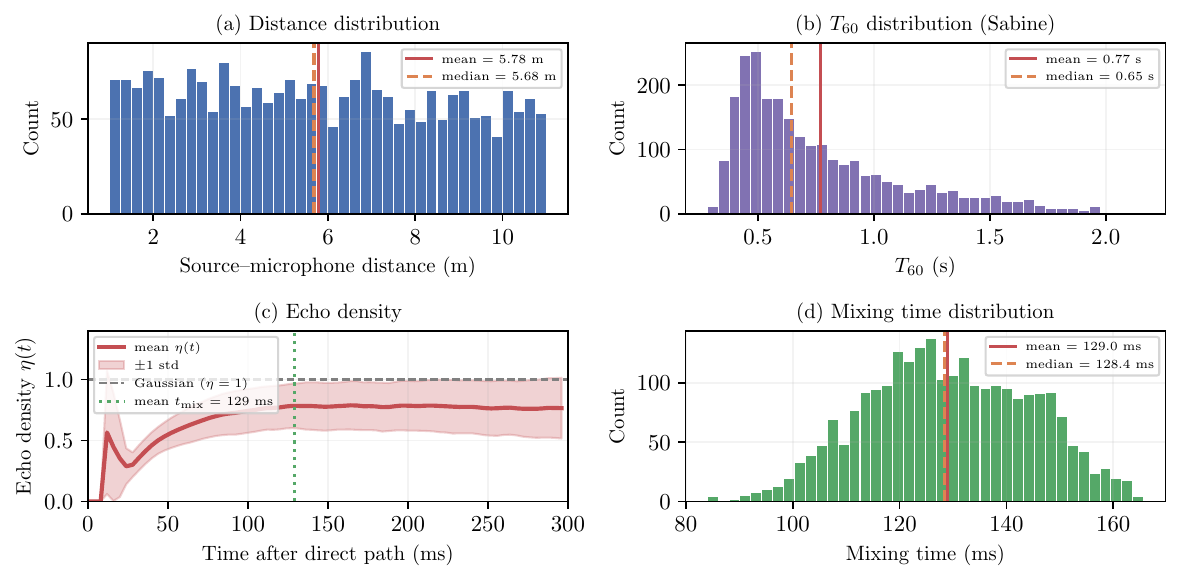}}
    \caption{%
    Dataset statistics (shared across all four scenarios, $N = 2500$ samples). (a)~Source-microphone distance distribution ($r \in [1, 11]$\,m, uniform sampling). (b)~Reverberation time $T_{60}$ estimated via Sabine's formula. (c)~Normalised echo density $\eta(t)$~\cite{abel2006}; the vertical line marks the mean mixing time $t_{\mathrm{mix}}$, confirming that $\eta \approx 1$ (Gaussian statistics) at the chosen early/late boundary. (d)~Distribution of the mixing time $t_{\mathrm{mix}}$.%
  }
    \label{fig:DatasetStats}
\end{figure*}

Anechoic speech recordings obtained from the EARS dataset~\cite{Richter_Interspeech_2024} are convolved with the simulated \acp{RIR} from \textit{pyroomacoustics}~\cite{Scheibler_ICASSP_2018}. Table~\ref{tab:sim} depicts the range of random parameters for data generation, following~\cite{Neri_WASPAA_2023, Neri_TASLP_2024}. The experiments include $2500$ audio files of $10$ s duration at $16$ kHz. The samples are randomly assigned to $5$ folds to assess the performance in a $5$-fold cross-validation fashion. By doing so, each cross-fold iteration assigns $1500$, $500$, and $500$ audios to training, validation, and testing sets, respectively. In this study, we impose the distribution of the distances to be uniform between $1$ and $11$ meters to avoid biasing the distance estimator with respect to short distances. This is done by generating room and source-receiver configurations that match the target distance. Detailed statistics of the synthetic dataset\footnote{Available at \href{https://github.com/michaelneri/audio-distance-estimation}{https://github.com/michaelneri/audio-distance-estimation}} are shown in Figure~\ref{fig:DatasetStats}.

\begin{table}[ht!]
\caption{Parameters for data generation.}
\label{tab:sim}
    \centering
        \adjustbox{max width=\textwidth}{%
    \begin{tabular}{cc}
    \hline
    \textbf{Parameter} & \textbf{Random ranges} \\
    \hline \hline
    Room width and length & $[3.0, 15.0]$ m \\
    Room height & $[2.0, 7.0]$ m \\
    \# of materials (wall, floor, ceiling)  & $13, 7, 8$ \\
    Source - receiver height & $[1.5, 2.2]$ m \\
    Source-to-surface distance & $> 0.5$ m \\
    Source-to-receiver distance & $> 1.0$ m \\
    \hline \hline
    \end{tabular}
}
\end{table}

\subsection{Room impulse response decomposition}
\label{sec:rir_decomposition}
 
A \ac{RIR} between a source at position $\mathbf{s}$ and a microphone at position $\mathbf{m}$ can be decomposed into three temporally non-overlapping regions:
\begin{equation}
  h(t) \;=\; \underbrace{h_{\mathrm{d}}(t)}_{\text{direct path}}
             \;+\; \underbrace{h_{\mathrm{e}}(t)}_{\text{early reflections}}
             \;+\; \underbrace{h_{\mathrm{l}}(t)}_{\text{late reverberation}},
  \label{eq:rir_decomposition}
\end{equation}
where each component occupies a successive time interval. The direct-path component $h_{\mathrm{d}}$ is a delayed and attenuated impulse whose time of arrival $\tau_{\mathrm{d}} = \lVert \mathbf{s} - \mathbf{m} \rVert / c$ is determined by the source-microphone distance and the speed of sound $c \approx \SI{343}{\metre\per\second}$. Early reflections $h_{\mathrm{e}}(t)$ consist of specular, sparse wall bounces that carry geometric information about the room. Late reverberation $h_{\mathrm{l}}(t)$ arises once reflections overlap sufficiently to form a diffuse, stochastic field.
 
The mixing time $t_{\mathrm{mix}}$ marks the onset of diffuse reverberation, i.e., the point beyond which the \ac{RIR} is well modeled by a Gaussian noise process~\cite{abel2006}. For each \ac{RIR}, we compute the normalised echo density $\eta(t)$ of Abel and Huang~\cite{abel2006}, and selecting $t = t_{\mathrm{mix}}$ such that $\eta(t) \approx 1$. Using the direct-path endpoint $t_{\mathrm{d}} = \tau_{\mathrm{d}} + \Delta t_{\mathrm{d}}$ (with $\Delta t_{\mathrm{d}} = \SI{2}{\milli\second}$ guard window) and the mixing time $t_{\mathrm{mix}}$ as boundaries, we derive four \ac{RIR} variants from each simulated impulse response by applying smooth half-cosine fade-outs and fade-ins ($\SI{5}{\milli\second}$ duration) at the relevant boundaries:
\begin{enumerate}
  \item \textbf{Full}. Raw \ac{RIR} $h(t)$;
  \item \textbf{Direct only}. $h(t)$ faded to zero after $t_{\mathrm{d}}$, retaining only the direct path;
  \item \textbf{No late}. $h(t)$ faded to zero after $t_{\mathrm{mix}}$, removing the late reverberant tail;
  \item \textbf{No early}. The direct path up to $t_{\mathrm{d}}$ plus the late reverberation from $t_{\mathrm{mix}}$ onward with early reflections suppressed.
\end{enumerate}

\begin{table}[t]
  \centering
  \caption{%
    Calibration scenarios. \checkmark: calibration cue available; \texttimes: unavailable (simulated via randomisation).%
  }
  \label{tab:calibration_scenarios}
  \small
  \adjustbox{max width=0.5\textwidth}{%
  \begin{tabular}{@{}lccl@{}}
    \toprule
    Scenario & Time-cal.\ & Level-cal.\ & Condition \\
    \midrule
    \textsc{Fully calibrated}
      & \checkmark & \checkmark
      & Synchronised capture, known source level \\
    \textsc{Time-calibrated}
      & \checkmark & \texttimes
      & Synchronised capture, unknown level \\
    \textsc{Level-calibrated}
      & \texttimes & \checkmark
      & Arbitrary onset, known source level \\
    \textsc{Uncalibrated}
      & \texttimes & \texttimes
      & Arbitrary onset, unknown level \\
    \bottomrule
  \end{tabular}
  }
\end{table}
These variants allow us to isolate the contribution of each temporal region to distance estimation. They can be visually inspected in Figure~\ref{fig:rir_variants}. By training and evaluating on every variant, we can determine whether the model relies primarily on direct-path attenuation, early-reflection patterns, or late reverberation statistics.

\subsection{Calibration scenarios}
The cues available for distance estimation depend on the degree of calibration of the recording setup. We identify two calibration dimensions: \textbf{Time calibration.}  Simulated \acp{RIR} introduce a leading silence because of synchronised capture between microphone and source, whose duration is equal to the propagation delay $\tau_{\mathrm{d}} = r/c$, encoding distance directly.  This cue is an artifact of the data simulation pipeline as in no real recordings the signal and the recording are perfectly time-calibrated. To remove it, we strip the natural delay from the convolved output and prepend a random silence $\delta \sim \mathcal{U}\{0,\, \lfloor r_{\max}/(c/f_s) \rfloor\}$. The \ac{RIR} itself is never modified. \textbf{Level calibration.}  When the source emission level is known, the received energy decreases with distance via the $1/r^2$ law, providing a direct amplitude cue. In an uncalibrated scenario, different speakers produce varying levels, which breaks this correlation. We simulate this by applying a random gain $G \sim \mathcal{U}(-6, 6)$\,dB to the anechoic speech before convolution.
 
Crossing the two calibration dimensions yields four scenarios (Table~\ref{tab:calibration_scenarios}). All four share the \emph{identical} set of rooms, source-microphone configurations, \acp{RIR}, talkers, and speech segments. By doing so, any performance difference is attributable solely to the available calibration information.

\subsection{Baseline model}

To estimate the speaker distance from single-channel recordings, we employ the state-of-the-art \ac{CRNN} used in~\cite{Neri_TASLP_2024, Neri_WASPAA_2023}. Within each dataset condition, the model is trained and evaluated on all four \ac{RIR} variants (full, direct-only, no-late, no-early), yielding a $4 \times 4$ results matrix that disentangles the effect of calibration scenarios from the contribution of each temporal region of the \ac{RIR}. The performance evaluation of the distance estimator is carried out by computing the \ac{MAE} $\mathcal{L}_1(y, \hat{y}) = |y-\hat{y}|$, where $y \in \mathbb{R^+}$ and $\hat{y} \in \mathbb{R^+}$ are the ground truth and predicted distances, respectively. In addition, following~\cite{Neri_TASLP_2024}, we also compute the relative \ac{MAE} $ r \mathcal{L}_1(y, \hat{y}) = \frac{\mathcal{L}_1(y, \hat{y})}{y} =  \frac{|y-\hat{y}|}{y}$.

\begin{figure*}[hbt!]
    \centering
    \centerline{\includegraphics[width=0.93\linewidth]{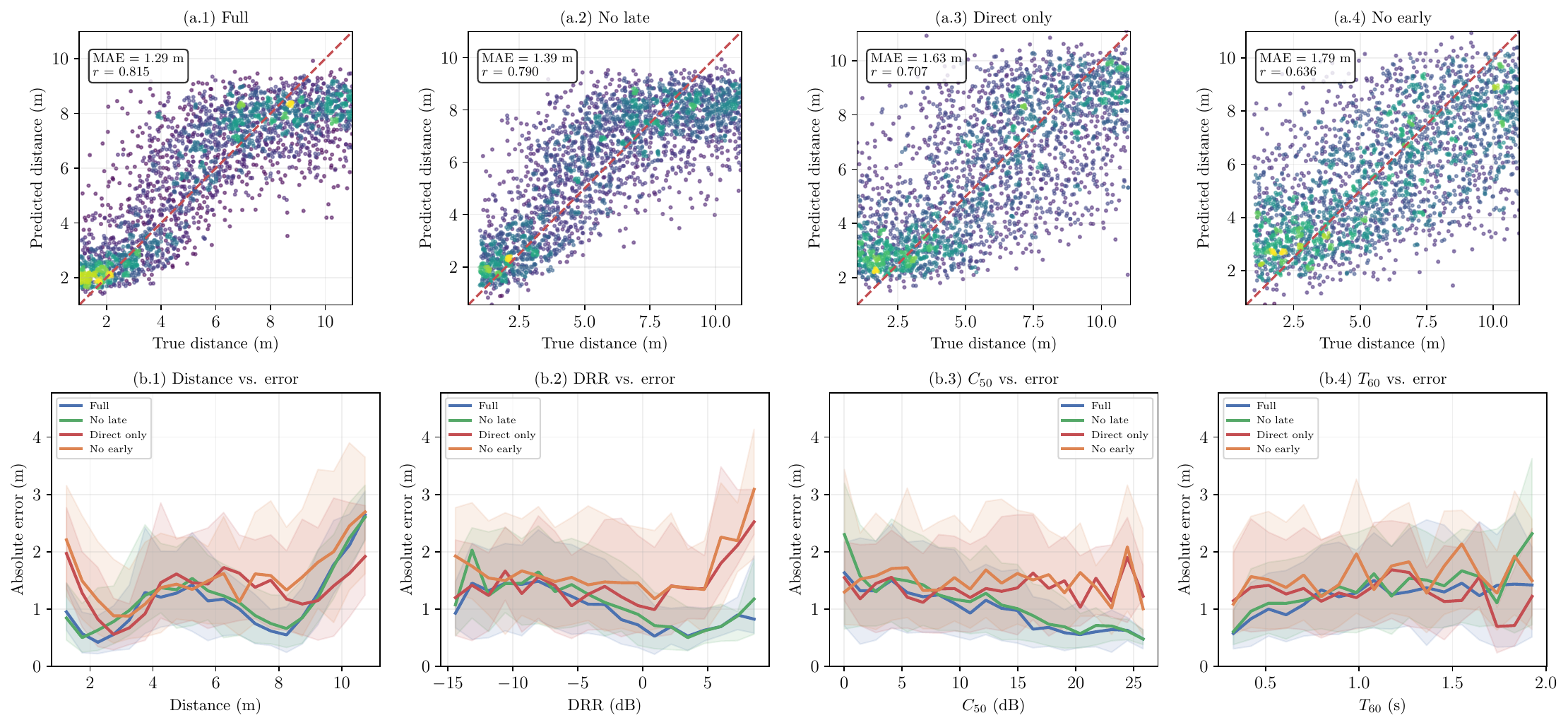}}
    \caption{Performance analysis under the uncalibrated scenario across all
    four \ac{RIR} variants.
    \textbf{Top row}~(a.1--a.4): predicted vs.\ true distance for
    each variant, with MAE and Pearson correlation $r$ reported.
    Point colour encodes local density.
    \textbf{Bottom row}~(b.1--b.4): median absolute error (solid
    lines) with interquartile range (shaded bands) as a function of
    source-microphone distance, \ac{DRR}, $C_{50}$, and $T_{60}$.
    }
    \label{fig:uncalibrated_analysis}
\end{figure*}

\begin{table*}[hbt!]
  \centering
  \caption{%
    \ac{MAE} with $95\%$ confidence intervals and relative MAE (\%) for each combination of  calibration scenario and \ac{RIR} variant. The best result per row is shown in \textbf{bold}. The \emph{Random} row reports the expected error of a model that has been trained with silent recordings exploiting only distance distribution, serving as a sanity check.%
  }
  \label{tab:results}
  \small
  \begin{tabular}{@{} l c c c c c c @{}}
    \toprule
    & & &
    \multicolumn{4}{c}{\textbf{RIR variant}} \\
    \cmidrule(l){4-7}
    \textbf{Scenario}
      & \textbf{Time}
      & \textbf{Level}
      & \textbf{Full}
      & \textbf{Direct}
      & \textbf{No late}
      & \textbf{No early} \\
    \midrule
    \multirow{2}{*}{Fully calibrated}
      & \multirow{2}{*}{\checkmark}
      & \multirow{2}{*}{\checkmark}
      & $0.15\;{\scriptstyle [0.14,\, 0.16]}$
      & $\mathbf{0.14}\;{\scriptstyle [0.13,\, 0.16]}$
      & $0.15\;{\scriptstyle [0.14,\, 0.16]}$
      & $0.15\;{\scriptstyle [0.13,\, 0.16]}$ \\
    & &
      & $(3.5\%)$
      & $\mathbf{(3.4\%)}$
      & $(3.4\%)$
      & $(3.5\%)$ \\
    \addlinespace
    \multirow{2}{*}{Time-calibrated}
      & \multirow{2}{*}{\checkmark}
      & \multirow{2}{*}{\texttimes}
      & $0.15\;{\scriptstyle [0.13,\, 0.17]}$
      & $\mathbf{0.14}\;{\scriptstyle [0.12,\, 0.15]}$
      & $0.16\;{\scriptstyle [0.14,\, 0.17]}$
      & $0.15\;{\scriptstyle [0.14,\, 0.16]}$ \\
    & &
      & $(3.5\%)$
      & $\mathbf{(3.2\%)}$
      & $(3.7\%)$
      & $(3.5\%)$ \\
    \midrule
    \multirow{2}{*}{Level-calibrated}
      & \multirow{2}{*}{\texttimes}
      & \multirow{2}{*}{\checkmark}
      & $\mathbf{1.29}\;{\scriptstyle [1.17,\, 1.42]}$
      & $1.58\;{\scriptstyle [1.52,\, 1.64]}$
      & $1.38\;{\scriptstyle [1.28,\, 1.48]}$
      & $1.79\;{\scriptstyle [1.65,\, 1.94]}$ \\
    & &
      & $\mathbf{(29.0\%)}$
      & $(38.3\%)$
      & $(31.0\%)$
      & $(45.3\%)$ \\
    \addlinespace
    \multirow{2}{*}{Uncalibrated}
      & \multirow{2}{*}{\texttimes}
      & \multirow{2}{*}{\texttimes}
      & $\mathbf{1.29}\;{\scriptstyle [1.17,\, 1.41]}$
      & $1.63\;{\scriptstyle [1.52,\, 1.73]}$
      & $1.39\;{\scriptstyle [1.28,\, 1.50]}$
      & $1.79\;{\scriptstyle [1.65,\, 1.94]}$ \\
    & &
      & $\mathbf{(29.4\%)}$
      & $(41.3\%)$
      & $(31.5\%)$
      & $(44.9\%)$ \\
    \midrule
    \multirow{2}{*}{Random}
      & \multirow{2}{*}{--}
      & \multirow{2}{*}{--}
      & \multicolumn{4}{c}{$2.49\;{\scriptstyle [2.43,\, 2.55]}$} \\
    & &
      & \multicolumn{4}{c}{$(72.2\%)$} \\
    \bottomrule
  \end{tabular}
\end{table*}

\section{Experimental Results}
\label{sec:exp}

In this section a study on which temporal component of the \ac{RIR} is used in each scenario is analyzed. Further analysis is carried out on the effect of propagation delay and source level for the speaker distance estimation task.

\subsection{Dependence on RIR components}
\label{sec:analysis_uncalibrated}

Figure~\ref{fig:uncalibrated_analysis} presents the performance of all four RIR variants under the uncalibrated scenario, which reflects the most common practical condition. The no-early variant, which retains the direct path and late tail but suppresses early reflections, performs worst, even below direct-only. This identifies early reflections as the single most informative component for distance estimation when calibration cues are unavailable. Moreover, the no-late variant nearly performs as the full version, indicating that the combination of direct sounds and early reflections are the most important ones for distance estimation.

The ribbon plots (b.1--b.4) confirm this across acoustic conditions. Error decreases with increasing \ac{DRR} and $C_{50}$ for the full and no-late variants, but the direct-only and no-early variants remain at a higher error level regardless,  indicating they lack the distance information.  At very low \ac{DRR} ($< -10$\,dB), all variants converge to similarly high error. Increasing $T_{60}$ mildly degrades all variants, suggesting that while reverberation provides distance cues, excessive reverberation smears the temporal structure that carries them.

\subsection{Effect of propagation delay and source level}
\label{sec:calibration_results}

Table~\ref{tab:results} also reveals why prior works~\cite{Neri_TASLP_2024, Neri_WASPAA_2023} on simulated data achieved centimeter-level accuracy: the time-calibrated rows show that the propagation delay alone yields
$\mathcal{L}_1 \approx 0.14$\,m regardless of \ac{RIR} content. Removing level calibration has negligible effect, confirming that the amplitude cue is redundant when timing is available. Without the propagation delay, $\mathcal{L}_1$ increases by an order of magnitude.  Level calibration again
provides almost no benefit, indicating that the $1/r$ amplitude cue is a weak predictor compared to reverberation-based features.

All conditions remain well above the random baseline
($\mathcal{L}_1 = 2.49$\,m), confirming that the model extracts meaningful acoustic information even in the fully uncalibrated scenario.

\section{Conclusions}
\label{sec:concl}
This work presented a systematic analysis of how individual \ac{RIR} components contribute to single-channel speaker distance estimation under different calibration conditions. When time calibration is available, the propagation delay alone yields \ac{MAE}~$\approx 0.14$\,m and all \ac{RIR} components become redundant. Level calibration provides negligible benefit in every condition. In the uncalibrated scenario, the model shifts to reverberation-based cues, with early reflections emerging as the most informative component: removing them degrades performance to \ac{MAE}~$= 1.79$\,m, worse than using the direct path alone ($1.63$\,m), while retaining them without the late tail ($1.39$\,m) nearly matches the full \ac{RIR} ($1.29$\,m). Further correlation analysis against \ac{DRR}, $C_{50}$, and $T_{60}$ confirms that estimation quality is based on the balance between early and reverberant energy, with excessive reverberation degrading all variants. Future work will investigate the robustness of these findings under additive noise conditions, extend the framework to moving speakers with time-varying source-microphone distances, and explore multi-microphone configurations where spatial cues may complement the reverberation-based features identified here.

\newpage

\bibliographystyle{IEEEtran}
\bibliography{refs}

\end{document}